%% file: paper.tex
\documentclass[conference]{IEEEtran}
\IEEEoverridecommandlockouts
\usepackage{cite}
\usepackage{amsmath,amssymb,amsfonts}
\usepackage{comment}
\usepackage{graphicx}
\usepackage{textcomp}
\usepackage{xcolor}
\usepackage{xspace}
\usepackage{listings}
\usepackage{hyperref}
\hypersetup{colorlinks,allcolors=black}
\def\BibTeX{{\rm B\kern-.05em{\sc i\kern-.025em b}\kern-.08em
    T\kern-.1667em\lower.7ex\hbox{E}\kern-.125emX}}

\usepackage{algorithm,algorithmicx}
\usepackage[noend]{algpseudocode}
\usepackage{subcaption}
\usepackage{comment}

\begin{document}

\title{Enabling Multi-threading in Heterogeneous Quantum-Classical Programming Models}

\author{
\IEEEauthorblockN{
Akihiro Hayashi\IEEEauthorrefmark{1}
Austin Adams\IEEEauthorrefmark{1}
Jeffrey Young\IEEEauthorrefmark{1}
Alexander McCaskey\IEEEauthorrefmark{2}
Eugene Dumitrescu\IEEEauthorrefmark{3}\\
Vivek Sarkar\IEEEauthorrefmark{1}
Thomas M.~Conte\IEEEauthorrefmark{1}
}
\IEEEauthorblockA{
\IEEEauthorrefmark{1}\textit{Georgia Institute of Technology}
\IEEEauthorrefmark{2}\textit{NVIDIA Corporation}
\IEEEauthorrefmark{3}\textit{Oak Ridge National Laboratory}
}
Email: \{ahayashi,aja,jyoung9,vsarkar,conte\}@gatech.edu, amccaskey@nvidia.com, dumitrescuef@ornl.gov
}

\maketitle

\input{sections/abstract}


\input{sections/introduction}

\input{sections/motivation}
\input{sections/qcor}
\input{sections/design}
\input{sections/impl}
\input{sections/eval}
\input{sections/discussions}
\input{sections/related}
\input{sections/conclusions}

\section*{Acknowledgement}
\noindent We acknowledge DOE ASCR funding under the Quantum Computing Application Teams program, FWP number ERKJ347. We also acknowledge support for this work from NSF planning grant \#2016666, ``Enabling Quantum Computer Science and Engineering''.

\bibliographystyle{IEEEtran}
\bibliography{biblio.bib}

\end{document}

%% file: sections/abstract.tex
\begin{abstract}
While quantum computers enable significant performance improvements for certain classes of applications, building a well-defined programming model has been a pressing issue. 
In this paper, we address some of the key limitations to realizing a generic heterogeneous parallel programming model for quantum-classical heterogeneous platforms. We discuss our experience in enabling user-level multi-threading in QCOR~\cite{10.1145/3380964} as well as challenges that need to be addressed for programming future quantum-classical systems. 

Specifically, we discuss our design and implementation of introducing C++-based parallel constructs to enable 1) parallel execution of a quantum kernel with \texttt{std::thread} and 2) asynchronous execution with \texttt{std::async}. To do so, we provide a detailed overview of the current implementation of the QCOR programming model and runtime, and discuss how we add 1) thread-safety to some of its user-facing API routines, and 2) increase parallelism in QCOR by removing data races that inhibit multi-threading so as to better utilize available computing resources. 

We also present preliminary performance results with the Quantum++~\cite{10.1371/journal.pone.0208073} back end on a single-node Ryzen9 3900X machine that has 12 physical cores (24 hardware threads) with 128GB of RAM. The results show that running two Bell kernels with 12 threads per kernel in parallel outperforms running the kernels one after the other each with 24 threads (1.63$\times$ improvement). In addition, we observe the same trend when running two Shor's algorthm kernels in parallel (1.22$\times$ faster than executing the kernels one after the other). Furthermore, the parallel version is better in terms of strong scalability.

We believe that our design, implementation, and results will open up an opportunity not only for 1) enabling quicker prototyping of parallel-aware quantum-classical algorithms on quantum circuit simulators in the short-term, but also for 2) realizing a generic parallel programming model for quantum-classical heterogeneous platforms in the long-term. 
\end{abstract}

%% file: sections/introduction.tex
\section{Introduction}
Quantum computing is a rapidly evolving field that leverages the laws of quantum mechanics for computation. Since near-term quantum computers are susceptible to significant levels of noise, a hybrid combination of classical computers and quantum computers, namely {\em quantum-classical} computers, is explored to mitigate noise while achieving orders-of-magnitude performance improvements for certain classes of applications. Such a hybrid combination can be viewed as one realization of heterogeneous computing where different types of processing elements, including special purpose accelerators, simultaneously and asynchronously work together.

\begin{figure}
  \centering
  \includegraphics[scale=0.27]{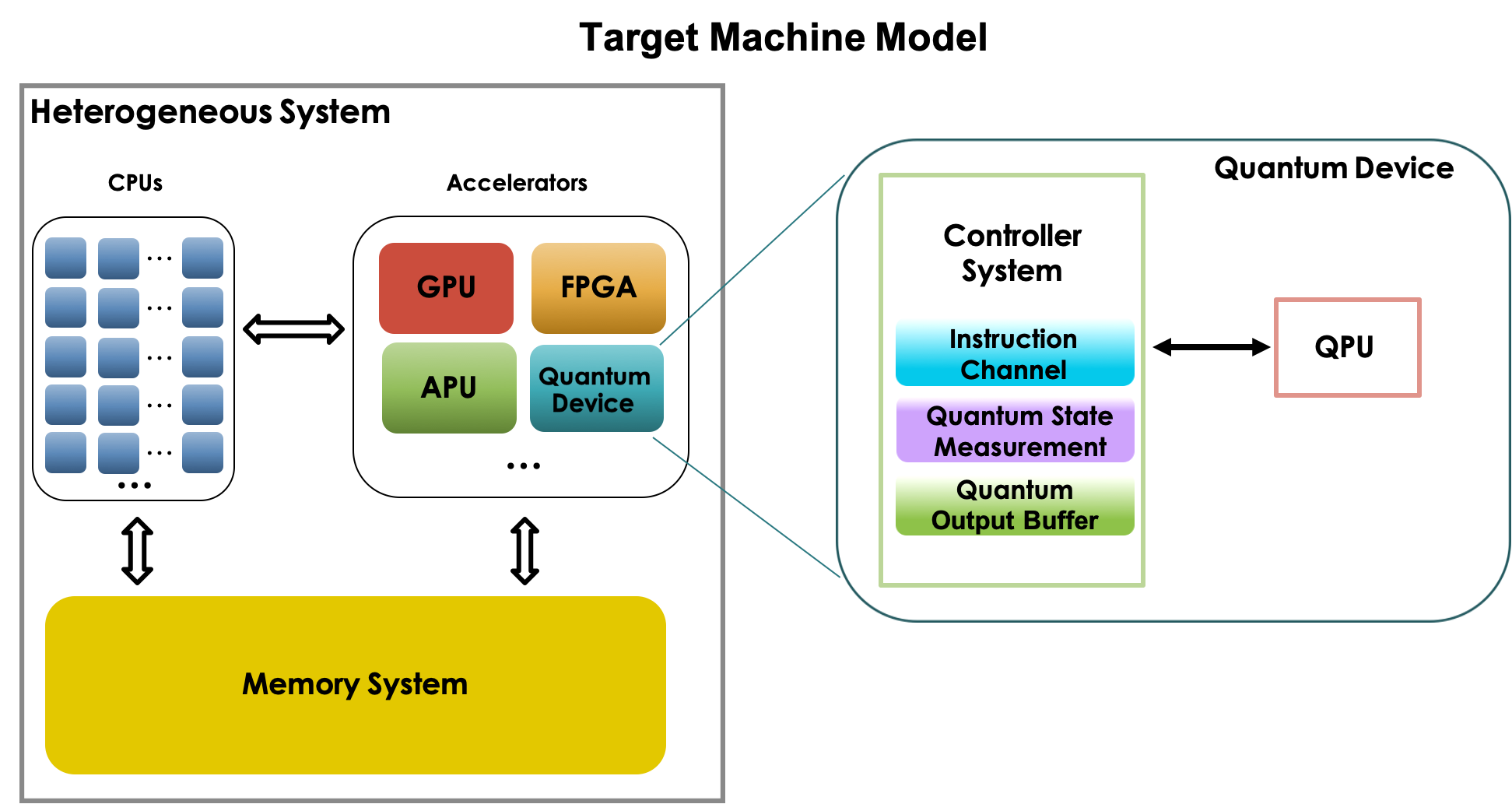}
  \caption{QCOR Machine Model~\cite{qcor-spec}}
  \label{fig:qcor-machine-model}
\end{figure}

QCOR~\cite{10.1145/3380964} is a programming system to realize such a heterogeneous quantum-classical model.  It is based on  the C++-based programming language and a compiler that is built on top of XACC~\cite{xacc_2020}  As shown in \autoref{fig:qcor-machine-model}, QCOR's target machine is a heterogeneous system where multiple CPUs (cores) are connected with quantum devices and other accelerators such as GPUs and FPGAs. 

To program quantum devices in QCOR, the user writes a \textit{quantum kernel} (i.e., a function that will be executed on a quantum device) in quantum computing domain-specific languages (DSLs), such as XACC's XASM or IBM's OpenQASM~\cite{Cross_2022}. Similar to other GPU-based heterogeneous programming models such as CUDA~\cite{cuda}, SYCL~\cite{sycl}, and OpenCL~\cite{opencl}, QCOR allows the user to write quantum kernels and CPU control code in the same program. This single-source programming model greatly facilitates quantum-classical programming.

However, one open research question for QCOR and other quantum DSLs is how to provide well-defined, user-level multi-threading support. Specifically, as the machine model in \autoref{fig:qcor-machine-model} implies, it is possible that multiple CPU cores might simultaneously utilize one or more quantum devices. Currently, there is no user-facing API-level support for multi-threading in quantum-classical programming models like QCOR and DSLs like OpenQASM, although it is typical to internally use multi-threading for accelerating quantum circuit simulations~\cite{10.1371/journal.pone.0208073,Suzuki2021qulacsfast ,Vincent2022jetfastquantum, 7877152}. 

In this paper, we explore the possibility of enabling user-level multi-threading in QCOR, which enables coarser grain parallelism in quantum-classical programming models. We believe this is an important step towards realizing an end-to-end heterogeneous programming system that can work on general heterogeneous platforms that include quantum computers. This work makes the following key contributions:

\begin{itemize}
    \item {Design and implementation of multi-threading support for a heterogeneous quantum-classical programming model.}
    \item {Discussion of scenarios and use cases where user-level multi-threading is beneficial for near-term quantum systems.}
    \item {A demonstration which shows that running two quantum kernels in parallel using $N/2$-threads for each kernel outperforms running the kernel one-by-one using $N$-threads, by factors of  1.22$\times$ to 1.63$\times$ for the evaluated kernels.}
 \end{itemize}

%% file: sections/motivation.tex
\section{Motivation}\label{sec:motivation}
This section highlights our motivation for enabling user-level multi-threading in quantum-classical computing by discussing potential parallelism in quantum-classical programs.

Let us use Shor's algorithm as a motivating example. In Algorithm~\ref{algo:shor}, \textsc{Shor} is a quantum-classical task that invokes the period-finding quantum kernel (\textsc{ShorKernel}) to estimate exponent $r$. Notice that \textsc{Shor} can be called multiple times until one or more (non-)trivial divisors are found or the entire search space is explored.

\begin{algorithm}[t]
\caption{Shor's Algorithm (Pseudocode)}\label{algo:shor}
\renewcommand{\algorithmicrequire}{\textbf{Input:}}
\renewcommand{\algorithmicensure}{\textbf{Output:}}
\algnewcommand\algorithmicforeach{\textbf{for each}}
\algdef{S}[FOR]{ForEach}[1]{\algorithmicforeach\ #1\ \algorithmicdo}
\begin{algorithmic}[1]
\Require{$N$: A natural number to be factorized.}
\Ensure{A non-trivial divisor(s) of $N$. }
\Procedure{Main}{N}
    \Repeat
    \State{$a \gets random(1, N);$} \Comment{$1 < a < N$}
    \State{$K \gets gcd(a, N);$}
    \If{$K == 1$}
        \State{\textsc{Shor}(N, a);}
    \Else{}
    \State {\Return {$K$}}
    \EndIf
    \Until{a divisor(s) is found or explored all}
\EndProcedure
\Procedure{Shor}{N, a}
  \For{$s = 1, ..., nShots$}
  \label{shotloop}        \State {$r_s \gets \textsc{ShorKernel}(N, a$)}
    \EndFor
    \State{$r \gets {r_1, ..., r_s}$} \Comment{Estimate $r$ from the measurements}
    \If{$r \mod 2 \equiv 1$ or $a^{r} \mod N \equiv -1$}
        \State{\Return{$\phi$};}
    \Else{}
        \State{\Return{$gcd(a^{r}/2 \pm 1, N)$};}
    \EndIf
\EndProcedure
\end{algorithmic}
\end{algorithm}

From the perspective of parallel processing, one possibility of parallelizing this algorithm is to run multiple instances of \textsc{Shor} in parallel. Furthermore, since it can require multiple shots to find $r$, it would be also possible to further parallelize the shot loop in \textsc{Shor} (Line~\autoref{shotloop}). Finally, if the \textsc{ShorKernel} is executed on a simulator, there is a massive amount of parallelism as in  \cite{10.1371/journal.pone.0208073,Suzuki2021qulacsfast ,Vincent2022jetfastquantum, 7877152}.
Algorithm~\autoref{algo:pshor} is a pseudo-parallel version of Algorithm~\autoref{algo:shor}. 
As in the X10 language~\cite{X10}, \textbf{async} represents parallel task creation and execution and \textbf{foreach}  represents parallel loop creation and execution.

\begin{figure*}
  \centering
  \includegraphics[scale=0.7]{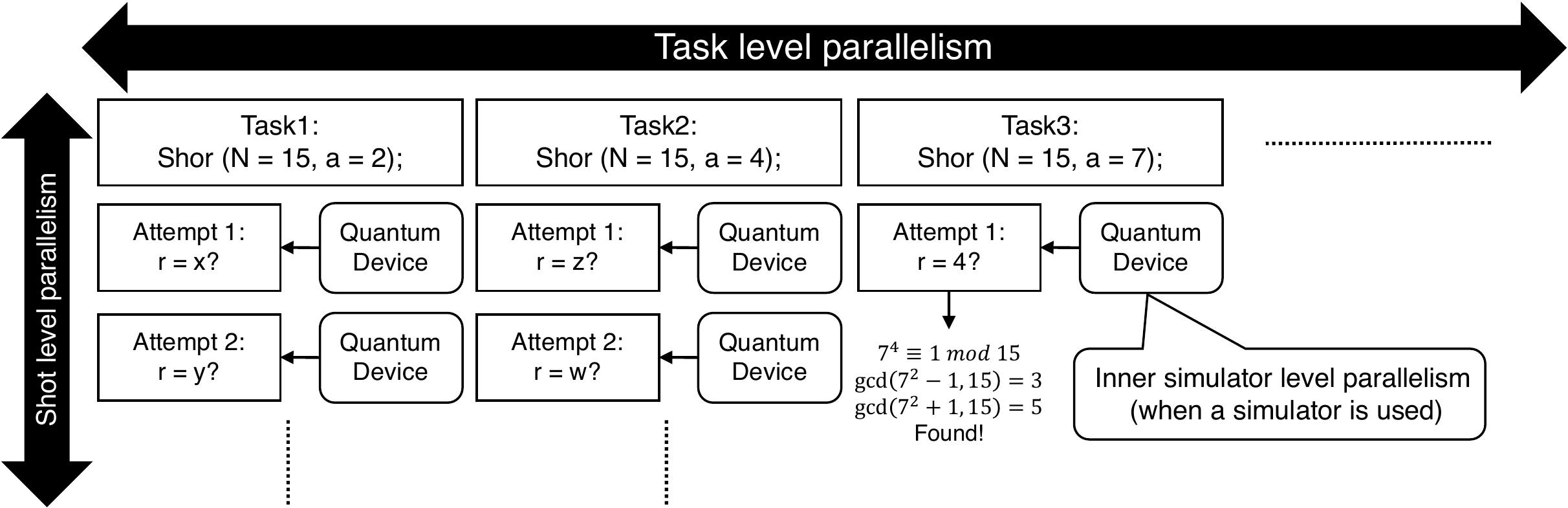}
  \caption{Multi-level parallelism in a quantum-classical program (Shor's algorithm).}
  \label{fig:parallel-shor}
\end{figure*}

\autoref{fig:parallel-shor} graphically illustrates the potential parallelism in Shor's algorithm across these three levels.
Based on what we discussed for Algorithm~\autoref{algo:pshor} and observe in \autoref{fig:parallel-shor}, we identify the following multiple levels of parallelism in quantum-classical programs:

\noindent
\textbf{Task level parallelism:} multiple independent classical tasks that can include quantum kernels are executed in parallel. 

\noindent
\textbf{Shot level parallelism:} multiple independent shots are executed in parallel.

\noindent
\textbf{Inner simulator level parallelism:} quantum simulators, including state vector and tensor network simulators such as \cite{10.1371/journal.pone.0208073,Suzuki2021qulacsfast ,Vincent2022jetfastquantum, 7877152}, are typically parallelized using OpenMP, CUDA, and the Eigen library to utilize a massive amount of parallelism on CPUs and/or GPUs. 

\begin{algorithm}[t]
\caption{Parallel Shor's Algorithm (Pseudocode)}\label{algo:pshor}
\renewcommand{\algorithmicrequire}{\textbf{Input:}}
\renewcommand{\algorithmicensure}{\textbf{Output:}}
\algnewcommand\algorithmicforeach{\textbf{for each}}
\algdef{S}[FOR]{ForEach}[1]{\algorithmicforeach\ #1\ \algorithmicdo}
\algrenewcommand\algorithmicforall{\textbf{foreach}}
\algdef{S}[FOR]{ForAll}[1]{\algorithmicforall\ #1\ \algorithmicdo}
\begin{algorithmic}[1]
\Require{$N$: A natural number to be factorized.}
\Ensure{A non-trivial divisor(s) of $N$. }
\Procedure{Main}{N}
    \Repeat
    \State{$a \gets random(1, N);$} \Comment{$1 < a < N$}
    \State{$K \gets gcd(a, N);$}
    \If{$K == 1$}
        \State{\textbf{async} \textsc{Shor}(N, a);}
    \Else{}
    \State {\Return {$K$}}
    \EndIf
    \Until{a divisor(s) is found or explored all}
\EndProcedure
\Procedure{Shor}{N, a}
    \ForAll{$s = 1, ..., nShots$}
        \State {$r_s \gets \textsc{ShorKernel}(N, a$)}
    \EndFor
    \State{$r \gets {r_1, ..., r_s}$} \Comment{Estimate $r$ from the measurements}
    \If{$r \mod 2 \equiv 1$ or $a^{r} \mod N \equiv -1$}
        \State{\Return{$\phi$};}
    \Else{}
        \State{\Return{$gcd(a^{r}/2 \pm 1, N)$};}
    \EndIf
\EndProcedure
\end{algorithmic}
\end{algorithm}

It is worth noting that the actual amount of available parallelism depends not only on algorithms but also on the simulated or physical quantum back ends that are targeted. One example would be when a user executes their program on a current-day single QPU system in which there would be limited parallelism due to the lack of additional physical hardware. However, in most cases, we believe that allowing the user to specify all available parallelism for a quantum-classical task will greatly enhance the performance and expressiveness of quantum-classical programs because there are plenty of computing resources (CPUs, GPUs, and FPGAs) that can accelerate the development of quantum-classical algorithms even on conventional systems.

Thus,  we believe that enabling user-level multi-threading in quantum-classical programming models will 1) accelerate the development of a quantum-classical algorithm, and 2) facilitate porting an existing heterogeneous algorithm to a quantum-classical one. It is also worth noting that the goal of this work is not optimizing and fine-tuning quantum-classical parallel programs for a specific target system. Instead,  we look to motivate and introduce concrete parallel programming constructs (\texttt{std::thread} and \texttt{std::async}) for quantum-classical programming models.

\begin{lstlisting}[float, caption=A 2-qubit Bell kernel implementation in QCOR, label=lst:bell, escapechar=|]
using namespace std;
// the Bell kernel
__qpu__ void bell(qreg q) { |\label{bell-start-seq}|
  using qcor::xasm;
  H(q[0]);
  CX(q[0], q[1]);
  for (int i = 0; i < q.size(); i++) {
    Measure(q[i]);
  }
}                           |\label{bell-end-seq}|   
int main(int argc, char **argv) {
  // Create two qubit registers, each size 2
  auto q = qalloc(2);        |\label{qalloc}|
  // Run the quantum kernel
  bell(q);                   |\label{bell-seq}|
  // dump the results
  q.print();                 |\label{bell-print-seq}|
}
\end{lstlisting}
\begin{lstlisting}[float, caption=An example output of the Bell kernel (1024 shots), label=lst:bell-output, escapechar=|]
"AcceleratorBuffer": {
  "name": "qrg_bmQBh",
  "size": 2,
  "Information": {},
  "Measurements": {
     "00": 513,
     "11": 511
   }
}
\end{lstlisting}

\begin{lstlisting}[float, caption=A VQE implementation in QCOR, label=lst:vqe, escapechar=|]
__qpu__ void ansatz(qreg q, double theta) {
  X(q[0]);
  Ry(q[1], theta);
  CX(q[1], q[0]);
}

int main(int argc, char **argv) {
  // Allocate 2 qubits
  auto q = qalloc(2);

  // Programmer needs to set
  // the number of variational params
  auto n_variational_params = 1;

  // Create the Deuteron Hamiltonian
  auto H = 5.907 - 2.1433 * X(0) * X(1) - 
           2.1433 * Y(0) * Y(1) + .21829 * Z(0) -
           6.125 * Z(1);

  // Create the ObjectiveFunction
  auto obj = createObjectiveFunction(ansatz, H, q, 
                                     n_variational_params,
                                     {{"gradient-strategy", 
                                     "central"}, 
                                     {"step", 1e-3}});

  // Create the Optimizer.
  auto opt = createOptimizer("nlopt",
                             {{"nlopt-optimizer", 
                               "l-bfgs"}});
  // Optimize
  auto [opt_val, opt_params] = opt->optimize(objective);
  std::cout << opt_val << std::endl;
}
\end{lstlisting}

%% file: sections/qcor.tex
\section{QCOR}

QCOR is a C++-based high-level quantum-classical programming model. One of the key features of QCOR is that the user can write both quantum and classical kernels and functions in the same code. This feature is not only analogous to existing heterogeneous programming models such as CUDA, OpenCL, and SYCL, but it also also provides a new programming model for heterogeneous quantum-classical computing programs that achieve hybrid quantum-classical workflows. As shown in the machine model in \autoref{fig:qcor-machine-model}, in theory, the user is free to leverage different kinds of processors (e.g., CPUs, GPUs, FPGAs, Quantum Devices) that could all be enabled through a QCOR-style programming model. 

\autoref{lst:bell} shows an example of QCOR program that executes the Bell kernel. First, on Line \autoref{qalloc}, the \texttt{qalloc} API is called to allocate 2-qubits. Then, the kernel written in XASM is invoked on Line \autoref{bell-seq}. Notice that the kernel is defined on Line \autoref{bell-start-seq} - \autoref{bell-end-seq}. After the kernel is invoked, the measurement results can be inspected by printing the content of the quantum register as shown on Line \autoref{bell-print-seq}. An example output of the QCOR program can be found in \autoref{lst:bell-output}.

In addition to the simple quantum circuit simulation above, for completeness, we would like to emphasize that QCOR is expressive enough to write a wide variety of quantum-classical algorithms such as the variational quantum eigensolver (VQE) and the Quantum Approximate Optimization Algorithm (QAOA). \autoref{lst:vqe} shows a VQE implementation in QCOR. Note that \texttt{createObjectiveFunction} and \texttt{createOptimizer} are built-in QCOR helper functions that facilitate creating and invoking a classical optimizer with a user-defined objective function with the Deuteron Hamiltonian and the ansatz kernel. More details can be found in \cite{10.1145/3380964, qcor-spec}.

%% file: sections/design.tex
\section{Design}
\begin{lstlisting}[float, caption=Simultaneously Launching two Bell kernels (\texttt{std::thread}), label=lst:bell-threaded, escapechar=|]
using namespace std;
// the bell kernel
__qpu__ void bell(qreg q) { |\label{bell-start}|
  using qcor::xasm;
  H(q[0]);
  CX(q[0], q[1]);
  for (int i = 0; i < q.size(); i++) {
    Measure(q[i]);
  }
}                           |\label{bell-end}|   
void foo() {
  // Create two qubit registers, each size 2
  auto q = qalloc(2);
  // Run the quantum kernel
  bell(q);
  // dump the results
  q.print();
}
int main(int argc, char **argv) {
  thread t0(foo); thread t1(foo);
  // Other classical/quantum work
  ...
  t0.join(); t1.join();
}
\end{lstlisting}

\subsection{Multi-threading Design Overview}
Since QCOR is primarily written in C++, we look to enable user-level multi-threading in QCOR in a way that is acceptable to both QCOR and C++ programmers. For QCOR programmers, our goal is to minimize modifications to the code required for enabling multi-threading. For C++ programmers, our goal is to provide a threading interface that is natural to use. To that end, we leverage C++'s standard threading constructs (\texttt{std::thread} and \texttt{std::async}). However, in terms of general applicability, our discussions should apply to other parallel programming systems for C++, such as OpenMP~\cite{openmp}, Kokkos~\cite{9485033}, and RAJA~\cite{10.1145/3293883.3302577}.

\begin{lstlisting}[float, caption=Asynchronously Launching the Bell kernel (\texttt{std::async}), label=lst:bell-async, escapechar=|]
using namespace std;
int main(int argc, char **argv) {
  std::future<int> f = async(launch::async, 
                        [=]() -> int { foo(); return 1; });
  // Other classical/quantum work
  ...
  //
  f.get();
}
\end{lstlisting}

Our current focus is on enabling coarse-grain parallelism to exploit the full capability of a CPU-QPU system. In one scenario, the user would like a one-to-one relation between a CPU and a QPU to simultaneously perform $N$ independent tasks, where $N$ is the number of CPU-QPU pairs. Another scenario might be a one-to-many/many-to-one relation between CPU(s) and QPU(s). It is worth noting that the QPU part is not necessarily a hardware QPU device. Since QCOR offers different backends, the QPU part can be a quantum circuit simulation on either a local machine or a cloud service and can also incorporate coarser tasks such as VQE.

\subsection{User-Facing API}

\subsubsection{std::thread}
\autoref{lst:bell-threaded} shows an example where two threads simultaneously run the Bell kernel using \texttt{thread}. The main function creates two threads (\texttt{t0} and \texttt{t1}), each of which executes the \texttt{foo} function. In the \texttt{foo} function, it first allocates 2-qubits using \texttt{qalloc}, then invokes the kernel written in XASM in Line \autoref{bell-start} - \autoref{bell-end}, and finally gets the results. This approach enables the user to overlap other work on the main thread with the two threads. Also, the main function can wait on each thread by calling \texttt{join()}.

\subsubsection{std::async} Another example (\autoref{lst:bell-async}) is asynchronous execution where the main function asynchronously launches the \texttt{foo()} function with \texttt{async}. Similar to the \texttt{thread} example, the user may want to overlap other work with the launched task. However, one interesting difference is that \texttt{async} returns a \texttt{future} object, which helps the user to check the status of the asynchronously launched task and take further action depending on the return value of the task (\texttt{get()}).

\subsection{Enabling Thread Safety}
Thread safety is usually attributed to a function/routine that can be safely invoked by multiple threads simultaneously. It is very common that thread safety is guaranteed in conventional heterogeneous programming models such as CUDA, OpenCL, and SYCL. For example, the SYCL specification~\cite{sycl2020-spec} describes this in the following manner:  ``\textit{SYCL guarantees that all the member functions and special member functions of the SYCL classes described are thread safe.}''

It is worth noting that enabling thread safety does not necessarily mean improving performance because it essentially prevents multiple threads from simultaneously accessing shared data. 
In this work, our first priority is to enable thread safety for QCOR's user-facing API. For portions where parallelization is important, we explore the possibility of increasing parallelism in \autoref{sec:impl}.

%% file: sections/impl.tex
\section{Implementation}\label{sec:impl}
\begin{lstlisting}[float, caption=Making \texttt{qalloc()} thread-safe with Mutex Lock, label=lst:mutex, escapechar=|]
mutex m;
qbit qalloc(const int n) {
  lock_guard<mutex> lock(m);
  ...
  allocated_buffers.insert({...}); |\label{line:mutex}|
  ...
}
\end{lstlisting}
\begin{lstlisting}[float, caption=How a QPU instance is declared and created, label=lst:qpp-org, escapechar=|]
 namespace xacc {
 namespace internal_compiler {
   // global variable
   std::shared_ptr<Accelerator> qpu = nullptr;
   ...
}}
// Getting an instance of qpp 
qpu = xacc::getAccelerator("qpp"); |\label{line:qpu}|
\end{lstlisting}
This section discusses how we enable user-level multi-threading in QCOR and XACC. 

Since the QCOR and XACC systems include over 200K lines of code written in modern C++, we focus on discussing a few common cases that can possibly inhibit user-level multi-threading. Essentially, these cases are focused on identifying potential sources of data races when multi-threading is added.

\subsection{Identifying sources of data races}\label{subsec:races}

\subsubsection{Global Variables}Global variables are the most common source of data races because these variables can be accessed simultaneously by multiple threads. The following is a global \texttt{std::map} object that is used to implement \texttt{qalloc()}.

\begin{footnotesize}
\begin{verbatim}
// global variable
map<string, shared_ptr<AcceleratorBuffer>> 
    allocated_buffers{};
\end{verbatim}    
\end{footnotesize}

Because \texttt{qalloc()} internally invokes \texttt{map}'s \texttt{insert()}, which is not thread-safe, concurrent invocations of \texttt{qalloc()} can be problematic. 

\subsubsection{Services}
QCOR depends on different software components provided by QCOR itself and XACC. Typically, \texttt{xacc::getService<T>(...)} is used to obtain a shared pointer to a specific service, namely \texttt{T} in this example. For services that do not derive \texttt{xacc::Cloneable}, the \texttt{xacc::getService<T>(...)} always returns a pointer to the same instance, which can be another source of a data race. The following is an example where a pointer to the \texttt{qpp} accelerator, a software simulator in QCOR/XACC (i.e., Quamtum++\cite{10.1371/journal.pone.0208073}), which is used to run the Bell kernel in \autoref{lst:bell-threaded} and \autoref{lst:bell-async}, is stored into \texttt{acc}.

\begin{footnotesize}
\begin{verbatim}
shared_ptr<Accelerator> acc; // a local variable
acc = xacc::getService<Accelerator>("qpp", ...);
\end{verbatim}
\end{footnotesize}

\begin{lstlisting}[float, caption=QPU Manager Implementation (Simplified), label=lst:qpumap, escapechar=|]
using namespace std;
class QPUManager {
    public:
        static QPUManager& getInstance() {
            static QPUManager instance; return instance;
        }
    private:
        QPUManager() {}
        map<thread::id, shared_ptr<Accelerator>> qpu_map;
    public:
        shared_ptr<Accelerator> getQPU();
        void setQPU(std::shared_ptr<Accelerator> _qpu);
};
\end{lstlisting}

Because \texttt{Accelerator} is not \texttt{Cloneable}, \texttt{getService<Accelerator>(...)} always returns the same \texttt{qpp} instance. This can cause a data collision since multiple threads can simultaneously register their gates to the same accelerator and can thus end up simulating an erroneous circuit.

\subsection{Implementation Details}
In general, we pursue the following two approaches to 
remove data races that inhibit multi-threading in QCOR and XACC: 1) enabling thread safety and 2) increasing parallelism. The former goal is achieved by adding safety to multi-threaded execution with mutex locks. The latter approach explores the possibility of leveraging multi-threading to accelerate user programs.

\subsubsection{Enabling thread-safety}
For enabling thread-safety, we leverage \texttt{std::mutex} or \texttt{std::recursive\_mutex} to enable mutual exclusions. For example, \autoref{lst:mutex} shows \texttt{qalloc()}, which has a non-thread-safe call in Line \autoref{line:mutex}. We first create a \texttt{mutex} object in the global scope, and then the object is used to create a critical section with \texttt{std::lock\_guard}.

\subsubsection{Increasing Parallelism}

For increasing parallelism, we use a quantum accelerator object (\texttt{qpu}) as a motivating example. In the original implementation, as shown in \autoref{lst:qpp-org}, the \texttt{qpu} object is declared as a global variable and is initialized by calling \texttt{xacc::getAccelerator()}, which internally calls \texttt{xacc::getService<Accelerator>()}. Thus, this example includes the two data race scenarios discussed above in \autoref{subsec:races}. 

We remove the data races by i) making \texttt{Accelerator} cloneable to create different instances every time \texttt{xacc::getAccelerator()} is called, and ii) providing a \texttt{map} that maps a current thread ID to the corresponding accelerator object, the latter of which is called \texttt{QPUManager}.

\autoref{lst:qpumap} shows a brief overview of \texttt{QPUManager}. \texttt{QPUManager} is implemented by using the singleton pattern and contains the setter and getter functions. The setter function takes the return variable of \texttt{xacc::getAccelerator()} and registers the accelerator instance along with a current thread id to the map. Similarly, the getter function returns a \texttt{qpu} instance that corresponds to a current thread.

\subsection{Current Implementation Status}
We have implemented these changes to enable thread-safety for QCOR and have created a pull request against the QCOR~\cite{qcor-pr}, QCOR\_SPEC~\cite{qcor-spec-pr}, and XACC~\cite{xacc-pr} repositories. For increasing multi-threaded parallelism, we have confirmed that the examples (\autoref{lst:bell-threaded} and \autoref{lst:bell-async}) and Shor's kernel work in a parallel fashion, and we plan to create another pull request to share that functionality.

One small limitation of our implementation is that the user needs to manually call \texttt{quantum::initialize()} API at the beginning of each thread so the runtime can register its thread ID to the \texttt{QPUManager}. In the future, we plan to create a compiler pass that automatically inserts this API call. Alternatively, we could provide \texttt{qcor::thread} and \texttt{qcor::async} wrappers for the original C++ constructs that internally call this initialization function.

%% file: sections/eval.tex
\section{Preliminary Performance Evaluation}\label{sec:eval}
This section presents the results of an empirical evaluation of our extended QCOR programming model and runtime implementation on a single-node platform to demonstrate its performance benefits.

\noindent
\textbf{Purposes:} 
The goal of our evaluation is two-fold:
\begin{enumerate}
\item to demonstrate that our extended QCOR programming model and runtime system with C++ threading model enables parallel quantum kernel execution.
\item to demonstrate that enabling parallel quantum kernel execution is beneficial in terms of performance.
\end{enumerate}
 
\noindent
\textbf{Platform:} We present the performance results on a single-node AMD server, which consists of a 12-core, 24-thread Ryzen9 3900X CPu running at 3.8GHz with 128GB of DRAM.

\noindent \textbf{Quantum Kernels:} 

We use the following quantum kernels written in XASM: 
\begin{enumerate}
	\item{\bf Bell Kernel:} The 2-qubit Bell kernel shown in \autoref{lst:bell-threaded}. The number of shots is 1024.
	\item{\bf Shor's Kernel:} The period-finding quantum kernel, which is based on \cite{10.5555/2011517.2011525}. The number of shots is 10.
\end{enumerate}

\begin{figure}
  \centering
  \includegraphics[scale=0.7]{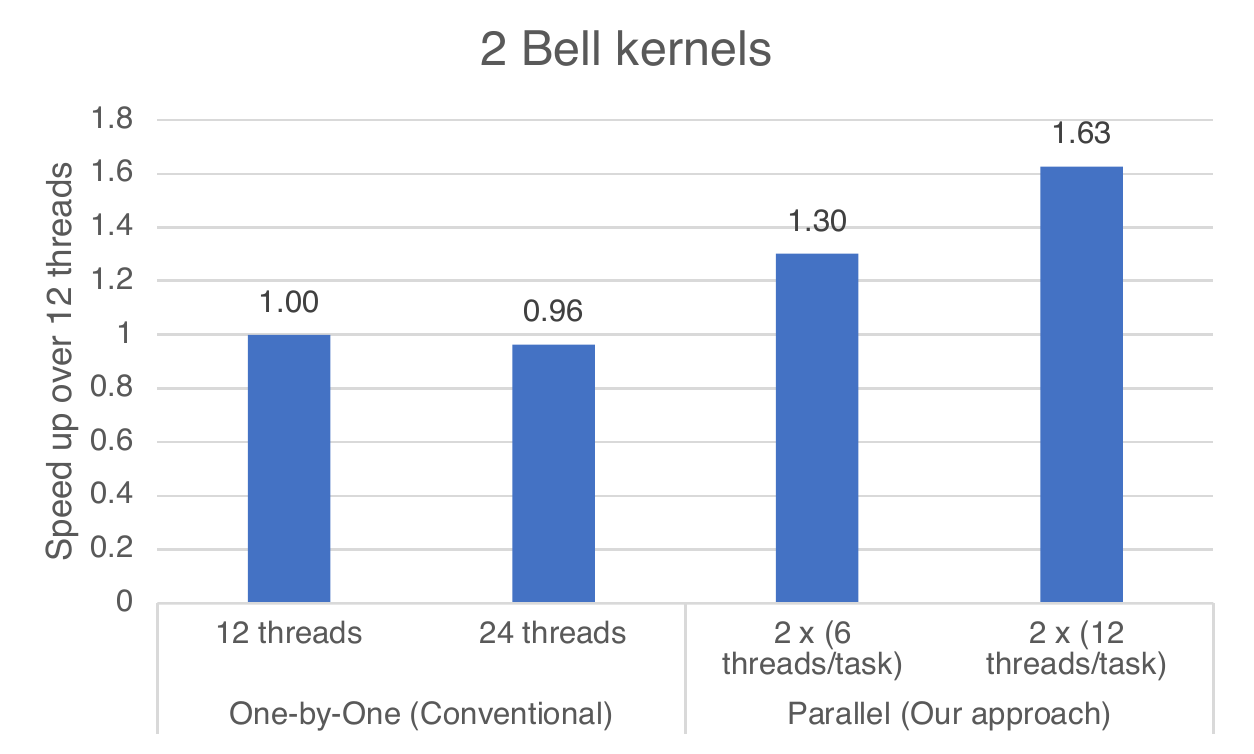}
  \caption{Bell Kernel}
  \label{fig:bell-results}
\end{figure}

\noindent \textbf{Experimental variants:}
For kernel simulations, we use the \texttt{QppAccelerator} backend in QCOR, which uses the Quantum++ library~\cite{10.1371/journal.pone.0208073}. 

We compare the following two variants in terms of performance:
\begin{enumerate}
	\item{\bf One-by-One (baseline, conventional):} Run the first kernel with $N$-threads and then run the second kernel with $N$-threads.
	\item{\bf Parallel:} Run the two kernels in parallel, each of which uses $N/2$-threads.
\end{enumerate}

Note that each kernel is executed on multiple physical cores/threads even in the baseline version because Quantum++ uses OpenMP~\cite{openmp}. For both variants, we appropriately set the \texttt{OMP\_NUM\_THREADS} parameter to specify the number of threads per kernel.  However, tuning this parameter for the best performance is beyond the scope of this paper.  Instead, our goal is to study scenarios where running multiple quantum kernels simultaneously could lead to performance benefits.  Finally, note that shot-level parallelism is not exploited in these versions. 

\subsection{Impact of Parallel Kernel Execution}
\begin{figure}
  \centering
  \includegraphics[scale=0.7]{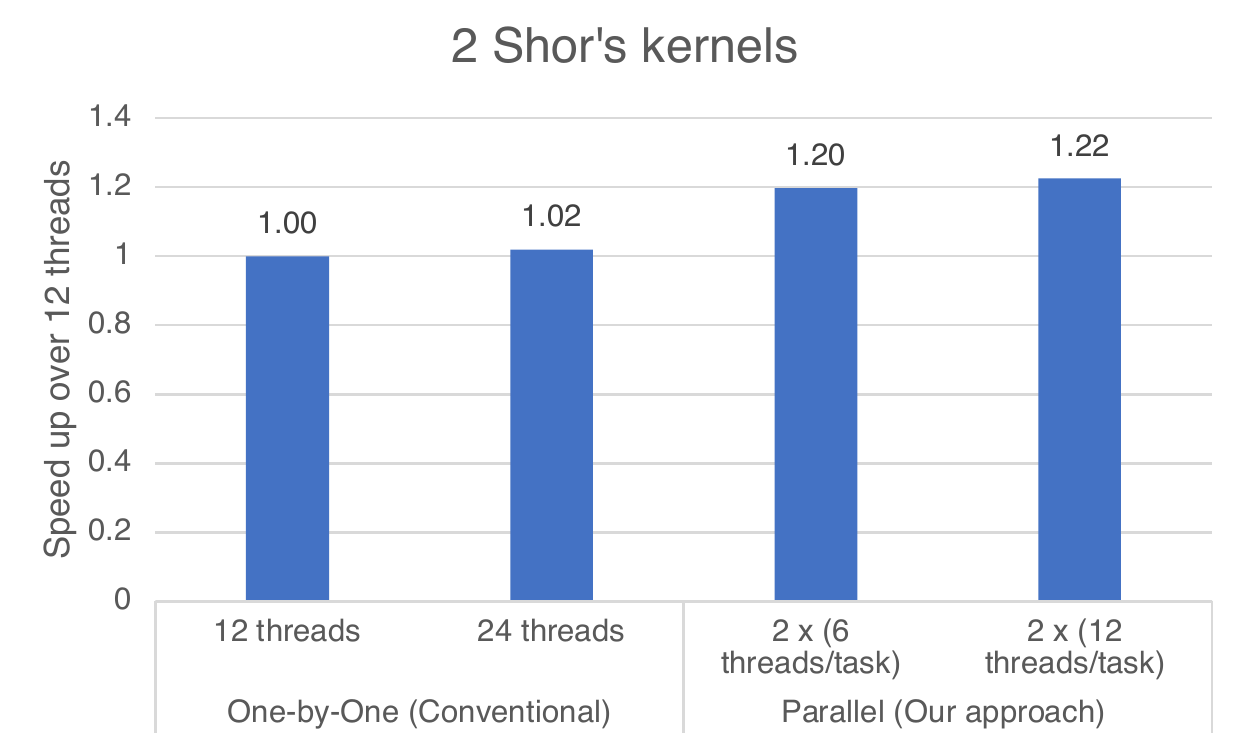}
  \caption{Shor's Kernel: \textsc{Shor}(N=15, a=2) and \textsc{Shor}(N=15, a=7) from Algorithm \autoref{algo:shor}}
  \label{fig:shor-results}
\end{figure}

\autoref{fig:bell-results} and \autoref{fig:shor-results} show relative performance improvements over the baseline execution (one-by-one execution with 12-threads). In one-by-one execution, increasing the number of threads from 12 threads to 24 threads does not improve performance. In contrast, parallel execution of the two kernels enables further performance improvements -i.e., 1.63$\times$ for the Bell kernel and 1.22$\times$  for Shor's kernel. Based on an analysis of this kernel using AMD $\mu$Prof, we observe that increasing the number of threads increases L1 data cache-related performance counter numbers such as \texttt{L1\_DC\_MISSES}. L1 misses get significantly worse, particularly when increasing the number of threads from 12 to 24, which is why the parallel 12 thread per task version is faster than the original version.

\subsection{Strong Scalability Study}
\begin{figure*}
  \centering
  \includegraphics[scale=0.7]{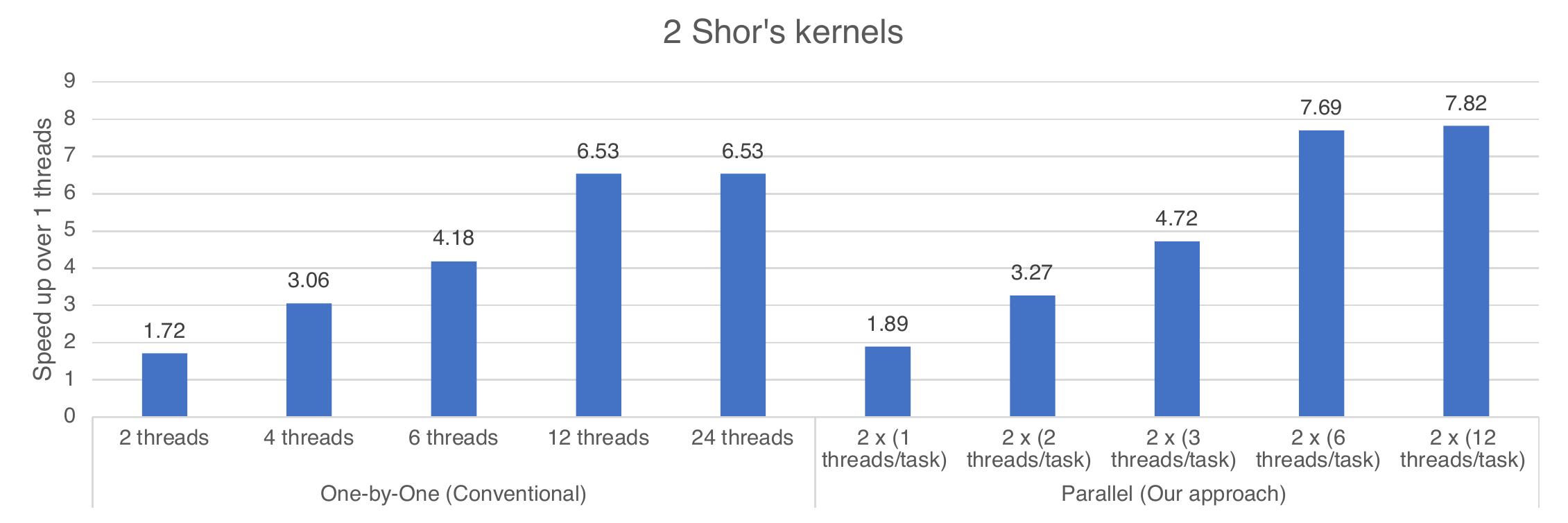}
  \caption{Scalability of the one-by-one and the parallel approaches: two \textsc{Shor}(N=7, a=2) from Algorithm \autoref{algo:shor}}
  \label{fig:shor-scalability-results}
\end{figure*}

\autoref{fig:shor-scalability-results} shows strong scalability of two Shor's kernels with the one-by-one and the parallel approaches. The numbers are relative performance improvements over the single-threaded one-by-one execution. While both approaches show good scalability, the parallel version always outperforms the baseline. 

%% file: sections/discussions.tex
\section{Discussion}\label{sec:discussions}
As shown in \autoref{sec:eval}, we demonstrated a scenario where running multiple kernels simultaneously is beneficial.  The goal of this section is to summarize difference application scenarios that we believe are good candidates for user-level multi-threading:

\noindent
\textbf{Shor's algorithm}: As we discussed in \autoref{sec:motivation}, suppose we factorize $N$ using Shor's algorithm, we can create $p$ parallel tasks with a random number $a_{p}$ s.t. $1 < a_{p} < N$ and $gcd(a_{p}, N) = 1$, each of which invokes Shor's kernel to estimate $r_{p}$ and checks if $r_{p}$ is even and $ $$a^{r_{p}} \mod N \equiv 1$ in parallel. Algorithm \autoref{algo:pshor} summarizes the parallel algorithm and \autoref{fig:shor-results} shows that running two Shor's kernels in parallel outperforms  one-by-one execution. We anticipate that the performance improvement will be more significant if CPUs with more cores and GPUs are used for simulating Shor's circuit.

\noindent
\textbf{VQE}: VQE~\cite{PeruzzoPhotonicVQE} optimizes a (Hamiltonian $H$) cost function over a parameterized manifold of quantum states $|\psi(\vec{\theta})\rangle=U(\vec{\theta})|\psi_0\rangle$ as $\underset{\vec{\theta}}{\text{min}}\langle \psi(\vec{\theta})|H|\psi(\vec{\theta})\rangle $. For QMA-hard Hamiltonians, dim$(\vec{\theta})$ is large but for many interesting models in physical sciences dim$(\vec{\theta})$ may scale (sub-)polynomially, in which case the optimization problem at hand may still be quite challenging. The pleasantly parallel nature of the optimization process can be utilized with multiple asynchronous quantum kernel instances minimizing over $\vec{\theta}$-space. 

\noindent
\textbf{Asynchronous Quantum JIT Compilation}: Shi et al.~\cite{10.1145/3297858.3304018} discusses a scenario where a GPU is used to compile and optimize quantum circuits, which can take several hours. With user-level multi-threading enabled, it is possible to avoid blocking computing resources by asynchronously offloading a compilation task onto a GPU and launching the compiled kernel on a QPU only when it is ready. 

\noindent
\textbf{Parallel Quantum-Classical Workflow}: As generalizations of different parallel execution scenarios discussed above, one can write an entire workflow in which different tasks run on different processing units including CPUs, QPUs, GPUs, and FPGAs. 

%% file: sections/related.tex
\section{Related Work}

While domain-specific languages (DSLs) for quantum computing significantly facilitate the development of quantum algorithms, many DSLs only focus on the kernel part and do not provide a system-wide programming model. We believe that such a system-wide programming model will become more important in quantum-classical computing because exploiting classical parallelism such as thread-level parallelism can improve end-to-end performance as discussed in \autoref{sec:eval}. Here, we briefly discuss existing programming models from the viewpoint of classical parallelism on non-quantum devices.

Qiskit~\cite{Qiskit} has been one of the most popular programming frameworks for quantum computing. However, it is not appropriate to directly map Qiskit programs to quantum-classical systems unless there is an AOT/JIT-level smart compiler that is aware of the underlying parallel hardware because the Global Interpreter Lock (GIL) may hinder Python-level multi-threaded execution.

Q\# is a programming language designed to express hybrid quantum-classical algorithms \cite{svore_q_2018}. Currently, there is no way to express the concept of threads in the Q\# language itself \cite{qsharp_spec}, nor in the Q\# standard library \cite{qsharp_stdlib}. Additionally, QIR (Quantum Intermediate Representation), a hybrid quantum-classical IR based on LLVM IR that is generated by the Q\# compiler front-end, does not explicitly guarantee thread-safety for any runtime functions \cite{qir_spec}. Indeed, the reference QIR runtime \cite{qir_runtime} may exhibit data races if used in multi-threaded code. It is worth noting that  QParallel~\cite{https://doi.org/10.48550/arxiv.2210.03680} allows the user to explicitly express parallelism in the quantum kernel part, not in the classical part.

Other newer platforms for hybrid quantum-classical computing have been proposed like NVIDIA's QODA~\cite{qoda}, which is designed for the simulation of quantum circuits with GPUs and QPUs. It is unclear what multi-threaded support model QODA uses as it is a proprietary product.

%% file: sections/conclusions.tex
\section{Conclusions and Future Work}

This paper explores the possibility of enabling user-level multi-threading in QCOR. We made enhancements to QCOR to support C++-based parallel and asynchronous execution of quantum kernels by 1) adding thread safety to QCOR API routines, and
2) increase parallelism by removing data races that inhibit multi-threading. 

Our preliminary results with the Bell and Shor's algorithm kernels show that enabling user-level multi-threading gives us performance improvements over the conventional baseline version in which each kernel is still executed by multiple threads, but is executed one-by-one. 

We believe this multi-threading design for heterogeneous quantum-classical programming models will open up an opportunity for rapidly prototyping and developing quantum-classical programs on conventional systems in the short-term. At the same time, we envision that this initial design would be a good starting point for longer-term explorations of heterogeneous programming systems for future quantum-classical systems.

In future work, we plan to run other quantum-classical tasks, such as VQE, with additional quantum simulation and physical back ends and also use different back ends to demonstrate where user-level multi-threading is most beneficial.

%% file: paper.bbl
\begin{thebibliography}{10}
\providecommand{\url}[1]{#1}
\csname url@samestyle\endcsname
\providecommand{\newblock}{\relax}
\providecommand{\bibinfo}[2]{#2}
\providecommand{\BIBentrySTDinterwordspacing}{\spaceskip=0pt\relax}
\providecommand{\BIBentryALTinterwordstretchfactor}{4}
\providecommand{\BIBentryALTinterwordspacing}{\spaceskip=\fontdimen2\font plus
\BIBentryALTinterwordstretchfactor\fontdimen3\font minus
  \fontdimen4\font\relax}
\providecommand{\BIBforeignlanguage}[2]{{%
\expandafter\ifx\csname l@#1\endcsname\relax
\typeout{** WARNING: IEEEtran.bst: No hyphenation pattern has been}%
\typeout{** loaded for the language `#1'. Using the pattern for}%
\typeout{** the default language instead.}%
\else
\language=\csname l@#1\endcsname
\fi
#2}}
\providecommand{\BIBdecl}{\relax}
\BIBdecl

\bibitem{10.1145/3380964}
\BIBentryALTinterwordspacing
T.~M. Mintz, A.~J. McCaskey, E.~F. Dumitrescu, S.~V. Moore, S.~Powers, and
  P.~Lougovski, ``Qcor: A language extension specification for the
  heterogeneous quantum-classical model of computation,'' \emph{J. Emerg.
  Technol. Comput. Syst.}, vol.~16, no.~2, mar 2020. [Online]. Available:
  \url{https://doi.org/10.1145/3380964}
\BIBentrySTDinterwordspacing

\bibitem{10.1371/journal.pone.0208073}
\BIBentryALTinterwordspacing
V.~Gheorghiu, ``Quantum++: A modern c++ quantum computing library,'' \emph{PLOS
  ONE}, vol.~13, no.~12, pp. 1--27, 12 2018. [Online]. Available:
  \url{https://doi.org/10.1371/journal.pone.0208073}
\BIBentrySTDinterwordspacing

\bibitem{qcor-spec}
{Oak Ridge National Laboratory Quantum Computing Institute}, ``{QCOR
  Specification},'' \url{https://github.com/ORNL-QCI/qcor\_spec}, 2022.

\bibitem{xacc_2020}
\BIBentryALTinterwordspacing
A.~J. McCaskey, D.~I. Lyakh, E.~F. Dumitrescu, S.~S. Powers, and T.~S. Humble,
  ``{XACC}: a system-level software infrastructure for heterogeneous
  quantum{\textendash}classical computing,'' \emph{Quantum Science and
  Technology}, vol.~5, no.~2, p. 024002, feb 2020. [Online]. Available:
  \url{https://doi.org/10.1088%2F2058-9565%2Fab6bf6}
\BIBentrySTDinterwordspacing

\bibitem{Cross_2022}
\BIBentryALTinterwordspacing
A.~Cross, A.~Javadi-Abhari, T.~Alexander, N.~de~Beaudrap, L.~S. Bishop,
  S.~Heidel, C.~A. Ryan, P.~Sivarajah, J.~Smolin, J.~M. Gambetta, and B.~R.
  Johnson, ``{OpenQASM}~3: A broader and deeper quantum assembly language,''
  \emph{{ACM} Transactions on Quantum Computing}, mar 2022. [Online].
  Available: \url{https://doi.org/10.1145%2F3505636}
\BIBentrySTDinterwordspacing

\bibitem{cuda}
\BIBentryALTinterwordspacing
J.~Nickolls, I.~Buck, M.~Garland, and K.~Skadron, ``Scalable parallel
  programming with cuda: Is cuda the parallel programming model that
  application developers have been waiting for?'' \emph{Queue}, vol.~6, no.~2,
  p. 40–53, mar 2008. [Online]. Available:
  \url{https://doi.org/10.1145/1365490.1365500}
\BIBentrySTDinterwordspacing

\bibitem{sycl}
{Khronos Group}, ``{SYCL Overview},'' \url{https://www.khronos.org/sycl/},
  2022.

\bibitem{opencl}
------, ``{OpenCL Overview},'' \url{https://www.khronos.org/opencl/}, 2022.

\bibitem{Suzuki2021qulacsfast}
\BIBentryALTinterwordspacing
Y.~Suzuki, Y.~Kawase, Y.~Masumura, Y.~Hiraga, M.~Nakadai, J.~Chen, K.~M.
  Nakanishi, K.~Mitarai, R.~Imai, S.~Tamiya, T.~Yamamoto, T.~Yan, T.~Kawakubo,
  Y.~O. Nakagawa, Y.~Ibe, Y.~Zhang, H.~Yamashita, H.~Yoshimura, A.~Hayashi, and
  K.~Fujii, ``Qulacs: a fast and versatile quantum circuit simulator for
  research purpose,'' \emph{{Quantum}}, vol.~5, p. 559, Oct. 2021. [Online].
  Available: \url{https://doi.org/10.22331/q-2021-10-06-559}
\BIBentrySTDinterwordspacing

\bibitem{Vincent2022jetfastquantum}
\BIBentryALTinterwordspacing
T.~Vincent, L.~J. O'Riordan, M.~Andrenkov, J.~Brown, N.~Killoran, H.~Qi, and
  I.~Dhand, ``Jet: {F}ast quantum circuit simulations with parallel task-based
  tensor-network contraction,'' \emph{{Quantum}}, vol.~6, p. 709, May 2022.
  [Online]. Available: \url{https://doi.org/10.22331/q-2022-05-09-709}
\BIBentrySTDinterwordspacing

\bibitem{7877152}
T.~Häner, D.~S. Steiger, M.~Smelyanskiy, and M.~Troyer, ``High performance
  emulation of quantum circuits,'' in \emph{SC '16: Proceedings of the
  International Conference for High Performance Computing, Networking, Storage
  and Analysis}, 2016, pp. 866--874.

\bibitem{X10}
\BIBentryALTinterwordspacing
P.~Charles, C.~Grothoff, V.~Saraswat, C.~Donawa, A.~Kielstra, K.~Ebcioglu,
  C.~von Praun, and V.~Sarkar, ``X10: An object-oriented approach to
  non-uniform cluster computing,'' vol.~40, no.~10, p. 519–538, oct 2005.
  [Online]. Available: \url{https://doi.org/10.1145/1103845.1094852}
\BIBentrySTDinterwordspacing

\bibitem{openmp}
\BIBentryALTinterwordspacing
L.~Dagum and R.~Menon, ``Openmp: An industry-standard api for shared-memory
  programming,'' \emph{IEEE Comput. Sci. Eng.}, vol.~5, no.~1, pp. 46--55, Jan.
  1998. [Online]. Available: \url{https://doi.org/10.1109/99.660313}
\BIBentrySTDinterwordspacing

\bibitem{9485033}
C.~R. Trott, D.~Lebrun-Grandié, D.~Arndt, J.~Ciesko, V.~Dang, N.~Ellingwood,
  R.~Gayatri, E.~Harvey, D.~S. Hollman, D.~Ibanez, N.~Liber, J.~Madsen,
  J.~Miles, D.~Poliakoff, A.~Powell, S.~Rajamanickam, M.~Simberg,
  D.~Sunderland, B.~Turcksin, and J.~Wilke, ``Kokkos 3: Programming model
  extensions for the exascale era,'' \emph{IEEE Transactions on Parallel and
  Distributed Systems}, vol.~33, no.~4, pp. 805--817, 2022.

\bibitem{10.1145/3293883.3302577}
\BIBentryALTinterwordspacing
D.~Beckingsale, R.~Hornung, T.~Scogland, and A.~Vargas, ``Performance portable
  c++ programming with raja,'' in \emph{Proceedings of the 24th Symposium on
  Principles and Practice of Parallel Programming}, ser. PPoPP '19.\hskip 1em
  plus 0.5em minus 0.4em\relax New York, NY, USA: Association for Computing
  Machinery, 2019, p. 455–456. [Online]. Available:
  \url{https://doi.org/10.1145/3293883.3302577}
\BIBentrySTDinterwordspacing

\bibitem{sycl2020-spec}
{Khronos Group}, ``{SYCL 2020 Specification},''
  \url{https://registry.khronos.org/SYCL/specs/sycl-2020/html/sycl-2020.html},
  2022.

\bibitem{qcor-pr}
{Hayashi, Akihiro}, ``{Initial Thread-Safe Implementation},''
  \url{https://github.com/qir-alliance/qcor/pull/157}, 2021.

\bibitem{qcor-spec-pr}
{Young, Jeffrey and Hayashi, Akihiro}, ``{Runtime section, multi-threaded
  support, reorganization of exec model},''
  \url{https://github.com/ORNL-QCI/qcor_spec/pull/9}, 2021.

\bibitem{xacc-pr}
{Hayashi, Akihiro}, ``{Initial Thread-Safe Implementation},''
  \url{https://github.com/eclipse/xacc/pull/455}, 2021.

\bibitem{10.5555/2011517.2011525}
S.~Beauregard, ``Circuit for shor's algorithm using 2n+3 qubits,''
  \emph{Quantum Info. Comput.}, vol.~3, no.~2, p. 175–185, mar 2003.

\bibitem{PeruzzoPhotonicVQE}
\BIBentryALTinterwordspacing
A.~Peruzzo, J.~McClean, P.~Shadbolt, M.-H. Yung, X.-Q. Zhou, P.~J. Love,
  A.~Aspuru-Guzik, and J.~L. O’Brien, ``A variational eigenvalue solver on a
  photonic quantum processor,'' \emph{Nature Communications}, vol.~5, no. 4213,
  Jul 2014. [Online]. Available: \url{http://dx.doi.org/10.1038/ncomms5213}
\BIBentrySTDinterwordspacing

\bibitem{10.1145/3297858.3304018}
\BIBentryALTinterwordspacing
Y.~Shi, N.~Leung, P.~Gokhale, Z.~Rossi, D.~I. Schuster, H.~Hoffmann, and F.~T.
  Chong, ``Optimized compilation of aggregated instructions for realistic
  quantum computers,'' in \emph{Proceedings of the Twenty-Fourth International
  Conference on Architectural Support for Programming Languages and Operating
  Systems}, ser. ASPLOS '19.\hskip 1em plus 0.5em minus 0.4em\relax New York,
  NY, USA: Association for Computing Machinery, 2019, p. 1031–1044. [Online].
  Available: \url{https://doi.org/10.1145/3297858.3304018}
\BIBentrySTDinterwordspacing

\bibitem{Qiskit}
A.~tA-v~et al., ``Qiskit: An open-source framework for quantum computing,''
  2021.

\bibitem{svore_q_2018}
\BIBentryALTinterwordspacing
K.~Svore, A.~Geller, M.~Troyer, J.~Azariah, C.~Granade, B.~Heim,
  V.~Kliuchnikov, M.~Mykhailova, A.~Paz, and M.~Roetteler, ``Q\#: {Enabling}
  {Scalable} {Quantum} {Computing} and {Development} with a {High}-level
  {DSL},'' in \emph{Proceedings of the {Real} {World} {Domain} {Specific}
  {Languages} {Workshop} 2018}.\hskip 1em plus 0.5em minus 0.4em\relax New
  York, NY, USA: Association for Computing Machinery, Feb. 2018, pp. 1--10.
  [Online]. Available: \url{https://doi.org/10.1145/3183895.3183901}
\BIBentrySTDinterwordspacing

\bibitem{qsharp_spec}
Microsoft, ``{Q\# Language Specification},''
  \url{https://github.com/microsoft/qsharp-language/tree/main/Specifications/Language},
  2022.

\bibitem{qsharp_stdlib}
------, ``{Microsoft Quantum Development Kit Libraries},''
  \url{https://github.com/microsoft/QuantumLibraries/}, 2022.

\bibitem{qir_spec}
{QIR Alliance}, ``{Quantum Intermediate Representation (QIR) Specification},''
  \url{https://github.com/qir-alliance/qir-spec/tree/main/specification}, 2022.

\bibitem{qir_runtime}
Microsoft, ``{The Native QIR Runtime},''
  \url{https://github.com/microsoft/qsharp-runtime/tree/main/src/Qir/Runtime},
  2022.

\bibitem{https://doi.org/10.48550/arxiv.2210.03680}
\BIBentryALTinterwordspacing
T.~Häner, V.~Kliuchnikov, M.~Roetteler, M.~Soeken, and A.~Vaschillo,
  ``Qparallel: Explicit parallelism for programming quantum computers,'' 2022.
  [Online]. Available: \url{https://arxiv.org/abs/2210.03680}
\BIBentrySTDinterwordspacing

\bibitem{qoda}
{NVIDIA}, ``{NVIDIA QODA: The Platform for Hybrid Quantum-Classical
  Computing},'' \url{https://developer.nvidia.com/qoda}, 2022.

\end{thebibliography}
